\documentclass[referee]{aa} 
\usepackage{epsfig}

\renewcommand\[{\begin{equation}}
\renewcommand\]{\end{equation}} 

\def\DPt{\widetilde{\Delta\Psi}}
\def\Dtau{\Delta (\tau)}

\def\M{{\cal M}}
\def\AM{A_{\rm M}}
\def\rhoz{\rho_0}
\def\rhot{\tilde\rho}
\def\rhoS{\rho_{\rm S}}
\def\phit{\tilde\phi}
\def\phiS{\phi_{\rm S}}
\def\phitz{\phit_0}
\def\phitu{\phit_1}
\def\phitd{\phit_2}

\def\rt{\tilde r}
\def\xt{\tilde x}
\def\yt{\tilde y}
\def\zt{\tilde z}
\def\Rt{\tilde R}
\def\lt{\tilde\ell}

\def\sigp{\sigma_{\varphi}}
\def\vphib{u_{\varphi}}

\def\vphibd{\vphib^2}

\def\vphidb{\overline{{\rm v}_{\varphi}^2}}

\def\vcirc{v_{\rm c}}
\def\vcirct{\tilde\vcirc}

\def\wrho{\widehat{\rho}}
\def\vn{u_{\rm n}}
\def\sign{\sigma_{\rm n}}
\def\sigpr{\sigma_{\rm p}}
\def\sigap{\sigma_{\rm a}}

\def\vpr{u_{\rm p}}

\def\Vpr{V_{\rm p}}
\def\vcx{u_x}
\def\vci{u_i}
\def\nv{{\bf n}}
\def\nci{n_i}
\def\ncj{n_j}
\def\sigij{\sigma_{ij}}
\def\siglos{\sigma_{\rm los}}

\def\xv{{\bf x}}

\def\Ylm{Y_{lm}}

\def\Dlm{D_{lm}}

\def\Sch{Schwarzschild$\;$}
\def\Jz{J_z}
\def\E{{\cal E}}
\def\Mp{M_{\rm p}}

\def\Mbh{M_{\rm BH}}
\def\mubh{\mu_{\rm BH}}
\def\Reff{R_{\rm e}}
 
\begin{document}

   \title{A simple method to construct exact
          density-potential pairs from a homeoidal expansion}

    \author{L. Ciotti\inst{1}
            \and
            G. Bertin\inst{2}}

    \offprints{L. Ciotti}

   \institute{Dipartimento di Astronomia, Universit\`a di Bologna,
              via Ranzani 1, I-40127 Bologna, Italy\\
              \email{luca.ciotti@unibo.it}
   \and      
              Dipartimento di Fisica, Universit\`a di Milano,
              via Celoria 16, I-20133 Milano, Italy\\
              \email{giuseppe.bertin@unimi.it}
              }

   \date{Accepted version, March 30, 2005}

\abstract{We start from a study of the density-potential relation for
classical homeoids in terms of an asymptotic expansion for small
deviations from spherical symmetry. We then show that such expansion
is a useful device that allows us to construct a variety of exact
density-potential pairs with spheroidal, toroidal, or triaxial shapes
for which the deviation from spherical symmetry is finite.  As
concrete analytical applications, we describe: (1) The construction of
a family of toroidal axisymmetric density-potential pairs one of which
is associated with a perfectly flat rotation curve (for a member of
this family, the supporting two-integral phase-space distribution
function is obtained in closed form); (2) The determination of the
aperture velocity dispersion in a wide class of two-integral
axisymmetric models not stratified on homeoids with central black
hole, which may be useful for the discussion of the dynamical
contributions to the characteristics of the Fundamental Plane of
early-type galaxies; and (3) For such class of models, the
construction of the $v/\sigma$-ellipticity relation, often considered
to assess the role of rotation in the structure of elliptical
galaxies.  \keywords{galaxies: kinematics and dynamics -- galaxies:
structure} } \titlerunning{Exact density-potential pairs} \maketitle
%

\section{Introduction}

For the discussion of many astrophysical problems where gravity is
important, a major difficulty is set by the potential theory. In
general, to calculate the gravitational potential associated with a
given density distribution one has to evaluate a three-dimensional
integral. Except for special circumstances, where a solution can be
found in terms of elementary functions, one has to resort to numerical
techniques and sophisticated tools, such as expansions in orthogonal
functions or integral transforms.

Under spherical symmetry, the density-potential relation can be
reduced to a one-dimensional integral, while for axisymmetric systems
one is left in general with a (usually non-trivial) two-dimensional
integral. As a result, the majority of explicit density-potential
pairs refers to spherical symmetry and only a handful of axially
symmetric pairs are known (e.g., see Binney \& Tremaine 1987,
hereafter BT87).  In few special cases (in particular, when a
density-potential pair is available in a suitable parametric form)
there exist systematic procedures to generate new non-trivial
density-potential pairs (e.g., see the case of Miyamoto \& Nagai 1975
and the related Satoh 1980 disks; see also Evans \& de Zeeuw 1992).

For non-axisymmetric systems the situation is worse. One class of
triaxial density distributions for which the potential can be
expressed in a tractable integral form is that of the stratified
homeoids, such as the Ferrers (1887) distributions (e.g., see
Pfenniger 1984, Lanzoni \& Ciotti 2003, hereafter LC03) and special
cases of the family considered by de Zeeuw \& Pfenniger
(1988). Additional explicit density-potential pairs are given by the
Evans (1994) models and by those constructed with the Kutuzov-Osipkov
(1980) method (see also Kutuzov 1998).

In this paper we draw attention to an elementary yet curious property
of the asymptotic expansion for small flattening of the homeoidal
potential quadrature formula.  Such expansion can be traced back to
the treatise on geodesy by Sir H. Jeffreys (1970, and references
therein; see also Hunter 1977). Recently, it has been applied to the
modeling of gaseous halos in clusters (Lee \& Suto 2003, 2004) and to
the study of the dynamics of elliptical galaxies (Muccione \& Ciotti
2003, 2004, hereafter MC03, MC04). The apparently unnoticed property
of the expansion is that it offers a device to construct, in a
systematic way, density-potential pairs with finite deviations from
spherical symmetry. In turn, these can be used to carry out a number
of calculations explicitly, thus allowing for a variety of interesting
applications to stellar dynamics.

The paper is organized as follows.  In Sect. 2 we present the basic
asymptotic expansions and we outline the method. In Sect. 3 we study
the Jeans equations for axisymmetric fluid models associated with
density distributions obtained from our method. In Sect. 4 we focus on
three significant applications. In Sect. 5 we make some concluding
remarks.  In Appendix A we address the issue of the relevant parameter
space. In Appendix B we clarify the relation between the homeoidal
expansion and the expansion in spherical harmonics.  In Appendix C we
summarize the intrinsic kinematical profiles of a family of
axisymmetric galaxy models with central black hole.
\begin{figure}[htbp]
\includegraphics[width=0.8\textwidth]{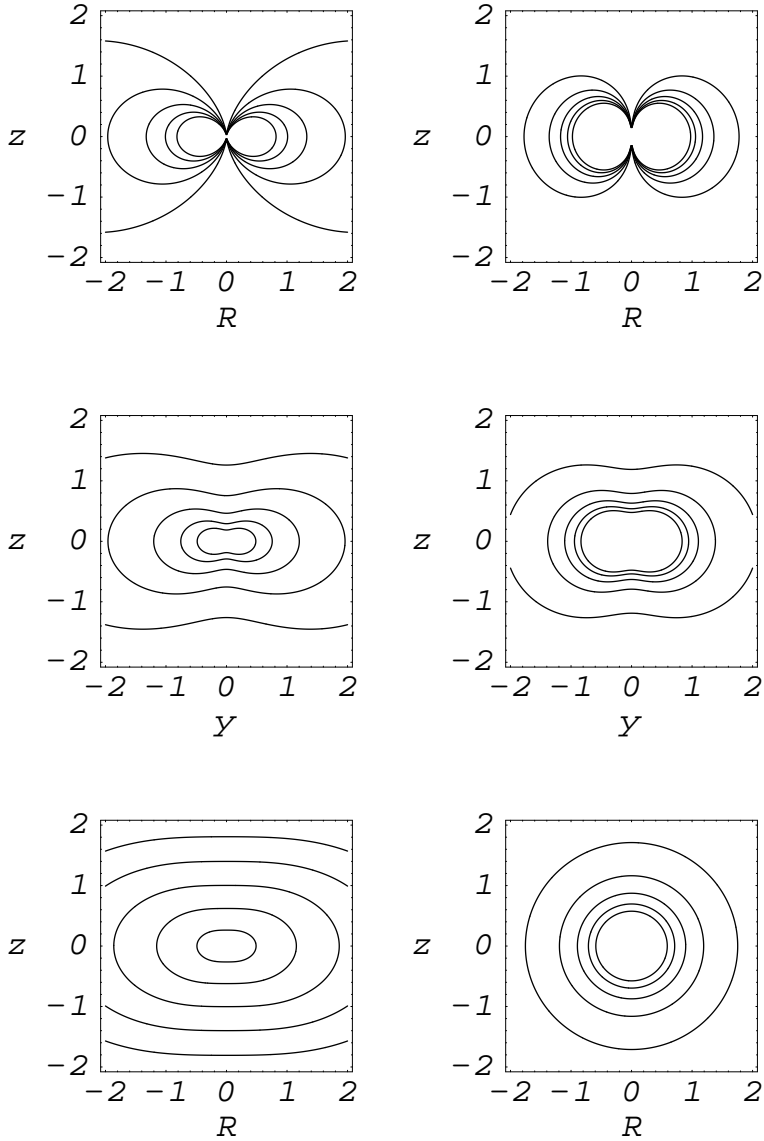}
\caption[]{Isodensity (top), constant projected density (middle), 
           and isopotential (bottom) contours (in arbitrary units) for
           the toroidal model described by Eq. (\ref{rhoT}) with
           $\alpha=3.1$ (left) and $\alpha=4.9$ (right).  The lower
           value of $\alpha$ is near the critical value for which the
           surface brightness is everywhere infinite, while the upper
           value is near the critical value for which the central mass
           diverges. The coordinates $R$, $y$, and $z$ are normalized
           to the scale-length $a$. Note how a strongly non-spherical
           density distribution, such as that represented on the top
           right, produces a nearly spherical potential (bottom
           right).}
\label{fig:torus}
\end{figure}

\section{The method}

We start by assuming a density distribution stratified on homeoidal
surfaces, $\rho(\xv)=\rhoz\rhot(m)$, where $\xv= (x,y,z)$, $\rhoz$ is
a density scale, and
\[
m^2={x^2\over a^2}+{y^2\over b^2}+{z^2\over c^2}=
{x^2\over a^2}+{y^2\over a^2(1-\epsilon)^2}+{z^2\over a^2(1-\eta)^2},
\]
with $a\geq b\geq c >0$, $b/a\equiv 1-\epsilon$, and $c/a\equiv
1-\eta$. The prolate case corresponds to $\epsilon=\eta <1$ and the
oblate case to $\epsilon =0$ and $0<\eta <1$.  In models of
astrophysical interest, the density $\rho$ may extend to infinity
(with finite or infinite mass) or be truncated. For simplicity, in
this paper we discuss only non-truncated density distributions.

It is well known (e.g., see Kellogg 1953, Chandrasekhar 1969) that the
(inner) potential associated with a homeoidal density distribution can
be written as
\[
\phi(\xv)=-\pi abc\rhoz G\int_0^{\infty}{\DPt (\xv;\epsilon,\eta)\over\Dtau}
           \;d\tau,
\label{poth}
\]
where $\Dtau=\sqrt{(a^2+\tau)(b^2+\tau)(c^2+\tau)}$, $\DPt
=2\int_{m(\xv;\tau)}^{\infty}\rhot(m)m\;dm$, and
$m^2(\xv;\tau)=x^2/(a^2+\tau)+y^2/(b^2+\tau)+z^2/(c^2+\tau)$.  If we
rescale density and potential to the quantities $\rhoz$ and $4\pi
G\rhoz a^2$, respectively, the Poisson equation for the dimensionless
density-potential pair $(\rhot ,\phit)$ becomes $\tilde\nabla^2\phit
=\rhot$ where $\tilde\nabla^2
=\partial^2/\partial\xt^2+\partial^2/\partial\yt^2+
\partial^2/\partial\zt^2$, $\xt\equiv x/a$, $\yt\equiv y/a$,
$\zt\equiv z/a$ and
\[
\phit=-(1-\epsilon)(1-\eta){a\over 4}
       \int_0^{\infty}{\DPt (\xv;\epsilon,\eta)\over\Dtau}\;d\tau. 
\label{pothd}
\]
This will be the starting point of our analysis.  We now expand the
density $\rhot (m)$ and the potential $\phit$ given in
Eq. (\ref{pothd}) up to the first significant order in the flattening
parameters and find:
\[
\rhot(m) =\rhot(\rt) +
          {\epsilon\yt^2 + \eta\zt^2 \over\rt}\rhot'(\rt)+
          O(\epsilon^2+\eta^2), 
\label{rhome}
\]
where $\rhot'\equiv d\rhot(m)/dm$ evaluated at $\epsilon=\eta=0$, and
\[
\phit =\phitz(\rt)+
       (\epsilon +\eta)\left[\phitu(\rt)-\phitz(\rt)\right]+
       (\epsilon\yt^2 + \eta\zt^2)\phitd(\rt)+
       O(\epsilon^2+\eta^2),
\label{pote}
\]
where
\[
\phit_i(\rt)=\cases{
\displaystyle{
-{1\over\rt}\int_0^{\rt}\rhot(m)m^2\;dm-
                        \int_{\rt}^{\infty}\rhot(m)m\;dm,\quad (i=0);}\cr
\displaystyle{
-{1\over 3\rt^3}\int_0^{\rt}\rhot(m)m^4\;dm-
 {1\over 3}\int_{\rt}^{\infty}\rhot(m)m\;dm,\quad (i=1);}\cr
\displaystyle{
{1\over\rt^5}\int_0^{\rt}\rhot(m)m^4\;dm,\quad (i=2).}\cr
}
\label{poteI}
\]
In the above expressions, the dimensionless spherical radius is
defined as $\rt\equiv r/a =\sqrt{\xt^2 +\yt^2 +\zt^2}$. Note that
$\phitz (\rt)$ is the potential associated with $\rhot (\rt)$, as can
be seen from its explicit expression. The infinite upper limit of
integration appearing in two integrals could be changed to a finite
constant if desired; this property is useful to deal with density
distributions for which the two integrals would diverge.  Similar
formulae can be obtained for the expansion at fixed total mass of
density distributions of finite total mass (e.g., see MC04,
Eq. [C.1]).

\subsection{Generating new families of non-spherical density-potential pairs}

By ordering arguments and by the linearity of the Poisson equation, it
follows that the truncation of Eqs. (\ref{rhome})-(\ref{pote}) to
first order in the flattening parameters produces {\it exact}
density-potential pairs independently of the value of $\epsilon$ and
$\eta$. In order to be physically acceptable, the density distribution
truncated to first order in $\epsilon$ and $\eta$ must be positive
definite (see Appendix A).  Density-potential pairs belonging to the
families thus constructed can be non-spherical at the ``non-linear''
level, because the values of the flattening parameters can differ
significantly from zero.  On the other hand, the truncated potential
is also the {\it approximate} potential for the density
$\rho=\rhoz\rhot(m)$ when $\epsilon\to 0$ and $\eta\to 0$.

Now we proceed one step further and use
Eqs. (\ref{rhome})-(\ref{pote}) to construct new density-potential
pairs, {\it independent of the ellipticities $\epsilon$ and $\eta$}
defining the homeoidal expansion. In fact, by the same ordering
argument used above, it is immediate to see that $\phitu
(\rt)-\phitz(\rt)+\zt^2\phitd (\rt)$ is the potential associated with
the ``density'' $\zt^2\rhot'(\rt)/\rt$. For radially declining density
distributions $\rhot(\rt)$, we can thus consider the (dimensionless)
pair
\[
\varrho=\zt^2{|\rhot'(\rt)|\over\rt};\quad \Phi=\phitz -\phitu -\zt^2\phitd.
\label{pairI}
\]
Two similar density-potential pairs can be then obtained trivially by
renaming variables, i.e., by replacing $\zt$ with $\yt$ or $\xt$ in
Eq. (\ref{pairI}).  We now combine the three density-potential pairs
linearly to get the following generally ``triaxial'' solution
\[
\varrho=(\alpha\xt^2+\beta\yt^2+\gamma\zt^2){|\rhot'(\rt)|\over\rt},
\label{pairIIr}
\]
\[
\Phi=(\alpha+\beta+\gamma)[\phitz (\rt)-\phitu (\rt)] -
({\alpha\xt^2+\beta\yt^2+\gamma\zt^2})\phitd (\rt).
\label{pairIIp}
\]
If we now set $\alpha=\beta=1$ and $\gamma=0$, we find a new
axisymmetric family characterized by toroidal geometry
\[
\varrho=\Rt^2{|\rhot'(\rt)|\over\rt};\quad 
        \Phi= 2(\phitz -\phitu)-\Rt^2\phitd.
\label{pairIII}
\]
Here $\Rt\equiv\sqrt{\xt^2+\yt^2}$ is the dimensionless cylindrical
radius. 

That Eqs. (\ref{pairI})-(\ref{pairIII}) indeed identify explicit
density-potential pairs can be checked {\it a posteriori} by deriving
the potential $\Phi$ from the density $\varrho$ by means of a standard
expansion in spherical harmonics (Appendix B).  Note however that, if
we started with a seemingly simpler idea, i.e., to transform the
density $\rhot (m)$ in spherical coordinates and then to expand it in
spherical harmonics, we would have discarded all the expansion terms
above the monopole as candidates for acceptable density distributions
because the related mass vanishes.  In contrast, the present method
works because, at least to first order, it is possible to isolate a
density term with {\it constant} (in our case, negative) sign over all
the space.

In the approach followed so far, we have considered the density
$\rho=\rhoz\rhot(m)$ as given: in this way, several density-potential
pairs can be easily obtained, for example by starting from well known
seeds for $\rhot (\rt)$, such as the density of the $\gamma$-models
(Dehnen 1993, Tremaine et al. 1994, see also Appendix A).  However,
the generated density profiles given in
Eqs. (\ref{pairI})-(\ref{pairIII}) remain largely out of control.

On the other hand, it is possible to specify the spherically symmetric
factor $\wrho\equiv |\rhot '(\rt)|/\rt$ in Eqs. (\ref{pairI}),
(\ref{pairIIr}), and (\ref{pairIII}): the potentials $\phitz$,
$\phitu$, and $\phitd$ required to complete the non--spherical
density-potential pair are recovered by inserting in
Eq. (\ref{poteI}) the {\it supporting density}
\[
\rhot(\rt)\equiv\int_{\rt}^{\infty}\wrho (t)t\;dt.
\label{suprho}
\]
The function $\wrho$ should go to zero sufficiently fast when
$r\to\infty$, in order for the integrals appearing in
Eq. (\ref{suprho}) to converge. We will use this approach to build the
special model presented in Sect. 4.1.

Another way to specify the supporting density $\rhot$ (and
so $\wrho$) is by imposing a given rotation curve in the equatorial
plane, that is a circular velocity $\vcirct (\Rt)$ defined from the
relation $\vcirct^2 (\Rt)\equiv\vcirc^2(R)/(4\pi G \rhoz a^2) =\Rt
\partial\Phi/\partial\Rt$ at $z=0$. In fact, by considering the
dimensionless potential $\Phi$ in Eq. (10) and by assigning the
function $\vcirct^2$, one gets and can solve the following
(inhomogeneous Euler) differential equation for $\rhot$:
\[
\Rt^4{d^2\rhot\over d\Rt^2}+
6\Rt^3{d\rhot\over d\Rt}+ 
2\Rt^2\rhot+{d\over d\Rt}
\left [{1\over\Rt}{d\over d\Rt}(\Rt^3\vcirct^2)\right ]=0.
\label{rotde}
\]


\section{The Jeans equations of stellar dynamics for axisymmetric 
         density-potential pairs (embedded, if desired, in a spherical 
         external field)}

In this Section, in view of applications to galactic dynamics, we
study the Jeans equations for axisymmetric density-potential pairs
either in the form of Eqs. (\ref{pairI}) and (\ref{pairIII}), or in
the axisymmetric cases of Eqs. (\ref{rhome})-(\ref{pote}).  In our
study we can also include the case in which the density distributions
mentioned above are added to a separate spherically symmetric density
distribution.

In all the axisymmetric cases described above, the pair $(\varrho,
\Phi)$ {belongs to the general family of density-potential 
pairs\footnote{In the self-gravitating case 
$A(r)\propto C''(r) + 2C'(r)/r + 4D(r)$ and $B(r)\propto D''(r) +6D'(r)/r$.} 
\[
\rho_{[R]}=A(r)+R^2 B(r),\quad \phi_{[R]}=C(r)+R^2D(r),
\label{pairg}
\]
where the subscript $[R]$ means that the pair is made explicit with
respect to $R$ by using (if required) the identity $z^2=r^2-R^2$.  The
possibility to switch to an alternative representation (i.e., from
$\rho_{[R]}$ and $\phi_{[R]}$ to $\rho_{[z]}$ and $\phi_{[z]}$) is
useful for the solution of the Jeans equations (when no subscript
appears, it means that either form can be considered). Note that for
$D(r)\propto 1/r^4$ the potential belongs to the class of St\"ackel
potential (St\"ackel 1890).

Let us assume that the density-potential pair defines the equilibrium
configuration of a stellar dynamical system for which the underlying
distribution function (DF) depends on the two classical integrals $E$
and $\Jz$ (star energy and $z$-component of the angular momentum, per
unit mass), so that the velocity dispersion tensor $\sigij$ is
diagonal and $\sigma^2\equiv\sigma_{RR}=\sigma_{zz}$; we will denote
by $\sigp^2$ the dispersion $\sigma_{\varphi\varphi}$. Along the
azimuthal direction, part of the kinetic energy can be stored in
systematic motions\footnote{A bar over a quantity means average over
phase-space velocities. In particular, $\vphib\equiv \overline{\rm
v_{\varphi}}=\int f {\rm v_{\varphi}}\,d^3{\bf v}/\rho$.}, so that
$\vphidb=\vphibd+\sigp^2$.

Given the form of the density-potential pair in Eq. (\ref{pairg}), 
the Jeans equations (e.g., see LC03, Eqs. [7]-[8]) can be easily
solved as
\[
\rho\sigma^2=\int_z^{\infty}\rho{\partial\phi\over\partial z'}\;dz'=
             \int_r^{\infty}\rho_{[R]}
             {\partial\phi_{[R]}\over\partial r'}\;dr',
\label{solJez}
\]
and 
\[
\rho\,(\vphidb-\sigma^2)=
                    R\left ({\partial\rho\sigma^2\over\partial R}+
                    \rho\,{\partial\phi\over\partial R}\right )=
                    {R^2\over r}
                    \left [{\partial (\rho\sigma^2)_{[z]}\over\partial r}+
                    \rho\,{\partial\phi_{[z]}\over\partial r}\right ].
\label{solJeR}
\]
Clearly an ambiguity remains as to the amount of support related to
systematic motions with respect to that of random motions in the
toroidal direction. A commonly used decomposition assumes
$\vphibd\equiv k^2(\vphidb-\sigma^2)$, and so
$\sigma_{\varphi}^2=\sigma^2+(1-k^2)(\vphidb-\sigma^2)$, where $k$ is
generally taken to be a given constant (Satoh 1980, but see Ciotti \&
Pellegrini 1996).  When $k=0$ the system is azimuthally supported only
by velocity dispersion, and no net rotation is present; in the
opposite limit, $k=1$, the velocity dispersion is isotropic. Of
course, the above procedure can be applied only when
$\vphidb-\sigma^2\geq 0$ everywhere.

The general expression of projection integrals for axisymmetric
systems can be found in LC03.  For simplicity, in this paper we focus
on the case when the system is observed edge-on (i.e. from a
line-of-sight where the contribution of rotation is expected to be
largest), even though for density distributions as in
Eq. (\ref{pairg}) explicit projection formulae can be obtained for any
inclination of the line-of-sight direction $\nv$.  Thus, we project
along the $x$ axis, so that $\nv =(1,0,0)$ and the projection plane is
the $(y,z)$ plane. The projected density is given by
$\Sigma=\int_{-\infty}^{\infty}\rho dx$, while the mass-weighted
projected streaming velocity field is given by
$\Sigma\vpr=\int_{-\infty}^{\infty}\rho\vn dx$, where
$\vn\equiv\vci\nci=\vcx=-\vphib\sin\varphi$ and $\sin\varphi =y/R$;
similarly, the projection of the square of the streaming velocity
field is $\Sigma\Vpr^2\equiv\int_{-\infty}^{\infty}\rho\vn^2 dx$.
Finally, the line-of-sight velocity dispersion field is
$\siglos^2=\sigpr^2+\Vpr^2-\vpr^2$, where $\Sigma\sigpr^2\equiv
\int_{-\infty}^{\infty}\rho\sign^2 dx$, and
$\sign^2\equiv\sigij\nci\ncj=\sigma_{xx}=\sigma^2+(1-k^2)(\vphidb
-\sigma^2)\sin^2\varphi$ (see Eqs. [B5] in LC03).

\subsection{The problem of the supporting distribution function}

At the end of this Section, we may ask whether it would be possible to
reconstruct the phase-space DF for the general family of densities
given in Eq. (\ref{pairIII}).  As is well known, the DF
\[
f=f_0\Jz^2 h(\E)\Theta(\E),
\label{df}
\]
where $\E=-v^2/2 +\Psi$ is the star binding energy per unit mass, and
$\Theta$ is the Heaviside step function, produces a mass density given
by
\[
\varrho={2^{7/2}\pi f_0\over 3} R^2\int_0^{\Psi}(\Psi- \E)^{3/2}h(\E) d\E,
\]
without streaming velocity (e.g., Hunter 1977; for a more general
family of distribution functions containing Eq. (\ref{df}) as a
special case, see also Ciotti, Bertin \& Londrillo 2004). Equation
(17) can be used to determine the two-integral DF associated with a
density profile of the form given by Eq. (\ref{pairIII}), in the
limiting case when the total potential $-\Psi$ is dominated by an {\it
external}, spherically symmetric potential $-\Psi_{\rm ext}(r)$ (so
that, in such approximation, we ignore the contribution to the
potential given by the field produced by $\varrho$ itself). In
particular, three different strategies can be considered. In the first
two cases $h(\E)$ is assigned and then Eq. (17) is used to determine
either the spherical factor $\wrho(r)$ for given $\Psi_{\rm ext}(r)$,
or, alternatively, $\Psi_{\rm ext}(r)$ for given $\wrho (r)$. In the
third case, the two functions $\Psi_{\rm ext}(r)$ and $\wrho(r)$ are
assigned, and $h(\E)$ is obtained by standard Abel inversion of the
function $\wrho [\Psi_{\rm ext}]$. Of course, in each of these cases
the appropriate physical requirements must be satisfied, namely the
mass density distribution associated with $-\Psi_{\rm ext}$ via the
Poisson equation and the function $h$ must be non-negative. Note that
if $d\wrho/ dr\leq 0$ then $r$ provides the natural parameter to
obtain the mapping $\Phi=\Phi(\varrho,\Rt^2)$ from
Eq. (\ref{pairIII}). This is an important step, because
$\varrho=\varrho(\Phi,\Rt^2)$ is the function needed to recover the
two-integral DF of axisymmetric systems (cf. the Fricke [1952] series
expansion, the Lynden-Bell [1962], the Dejonghe [1986], and the Hunter
[1975] integral transforms, and finally the Hunter \& Qian [1993]
integral contour method; see also Sect. 4.1.2).

\section{Three applications}

Following the normalization introduced at the beginning of Sect. 2, in
this Section and in Appendix C, the density $\varrho$, the relevant
radii, and the potential $\Phi$ are meant to be normalized to $\rhoz$,
to the scale-length $a$, and to $4\pi G\rhoz a^2$, respectively.  It
follows that masses, projected densities, and velocities are
normalized to $\rhoz a^3$, $\rhoz a$, and $2 a\sqrt{\pi G\rhoz}$; for
simplicity, we omit the tilde above these quantities.

\subsection{Power-law tori}

In this subsection we describe a toroidal stellar system belonging to
the family associated with Eq. (\ref{pairIII}), for which the general
methods outlined in Sect. 3 allow us to obtain explicitly the relevant
kinematical profiles (and their projections on the plane of the sky),
even in the presence of a central black hole. For simplicity, we focus
here on the fully self-gravitating case, i.e. on the case when no
black hole is present. The simplest model is that of a power-law
torus, defined by the (normalized) density distribution
\[
\varrho={\Rt^2\over\rt^{\alpha}},\quad (\alpha >0).
\label{rhoT}
\]
The total mass of the model is infinite, independently of the value of
$\alpha$, an unpleasant property shared with spherically symmetric
(non-truncated) power-law models. The central mass is finite for
$\alpha <5$.  According to Eq. (\ref{suprho}) the supporting density
is $\rhot =\rt^{2-\alpha}/(\alpha-2)$ (for $\alpha > 2$), and so from
Eqs. (\ref{pairIII}) and (\ref{poteI})
\[
\Phi=\cases{\displaystyle{
            -{\rt^{2-\alpha}\over (\alpha-2)(7-\alpha)}\left [ 
            {4\rt^2\over (\alpha-4)(5-\alpha)}+\Rt^2\right],
            \quad (\alpha\neq 4),}\cr
            \displaystyle{{1\over 3} \left(2\ln\rt -
            {1\over 2}{\Rt^2\over\rt^2}\right),\quad (\alpha =4).}
}
\label{potT}
\]
For $4<\alpha <5$ the gravitational potential vanishes for
$r\to\infty$ and diverges at the origin: thus, only orbits with
negative total energy are bound. For $2<\alpha \leq 4$ an upper
truncation must be applied to some integrals appearing in Eq.
(\ref{poteI}); the additive constant resulting in Eq. (\ref{potT}) has
been set equal to zero, and the potential vanishes at the
origin. Finally, for $\alpha =4$ the potential diverges both
at $r=0$ and $r=\infty$ and so orbits are bound independently of their
energy.

For $2<\alpha <5$ and $\alpha\neq 0$ the circular velocity in the
equatorial plane is
\[
\vcirc^2= {(-\alpha^2+9\alpha-16)\Rt^{4-\alpha}\over 
         (\alpha-2)(5-\alpha)(7-\alpha)},
\label{vcT}
\]
while $\vcirc^2=2/3$ for $\alpha=4$.  Curiously, the circular velocity
vanishes for $\alpha=\alpha_n\equiv 9/2-\sqrt{17}/2\approx 2.44$. In
fact, for $2<\alpha <\alpha_n$ Eq. (\ref{vcT}) would predict a {\it
negative} square circular velocity, in the sense that in this range of
$\alpha$ the radial force field in the equatorial plane, is directed
{\it outward}.  For $2<\alpha<5$, Eqs. (\ref{solJez})-(\ref{solJeR})
give
\[
\varrho\sigma^2={\Rt^2\,\rt^{2(1-\alpha)}\over 7-\alpha}\left[
                 {2\rt^2\over (\alpha-2)^2(5-\alpha)}+
                 {\Rt^2\over 2(\alpha-1)}\right ],
\label{sigT}
\]
and
\[
\varrho\,(\vphidb-\sigma^2)={2\Rt^2\,\rt^{2(1-\alpha)}\over 
                              (\alpha-2)(7-\alpha)}\left[ {2\rt^2\over
                              (\alpha-2)(5-\alpha)}-
                              {\Rt^2\over\alpha-1}\right ];
\label{vazT}
\]
since the r.h.s. of Eq. (\ref{vazT}) is positive, the Satoh
decomposition can be applied.

In the isotropic case ($k=1$), Eq. (\ref{vazT}) gives the square of
the streaming velocity $\vphibd$; in this case in the equatorial plane
(where $r=R$), the ratios $\sigma^2/\vcirc^2$, $\vphidb/\vcirc^2$, and
$\vphibd/\vcirc^2$ are constant.  The isotropic case can also be
interpreted as the description of a toroidal fluid structure for which
the velocity $\vphib$ is not constant over cylinders (corresponding to
the fact that the system is baroclinic; e.g., see Tassoul 1978): the
asymmetric drift $\vcirc^2 -\vphibd$ on the equatorial plane presented
by all the models scales as $R^{4-\alpha}$, vanishes for $\alpha=3$,
and is negative for lower values of $\alpha$.  Along the $z$ axis
(where $R=0$), according to Eq. (\ref{sigT}) the pressure vanishes,
and so does the quantity $\varrho(\vphidb-\sigma^2)$: while this
latter behavior is a common property of two-integral axisymmetric
systems (e.g., see Ciotti \& Pellegrini 1996, LC03), the former is a
peculiarity of the present toroidal model.  Along the major axis in
the edge-on projection plane (i.e., along the $y$ axis, at $z=0$),
$\vpr\propto k\yt^{2-\alpha/2}$, while
$\siglos^2=\sigpr^2+\Vpr^2-\vpr^2\propto\yt^{4-\alpha}$, where the
proportionality constants are simple functions of $\alpha$ that can be
easily computed for $\alpha >7/2$; below this value of $\alpha$ the
projected velocity dispersion diverges.  The properties of two
toroidal models associated with Eq. (\ref{rhoT}) are illustrated in
Fig. 1.

\subsubsection{The torus with flat rotation curve and the ``neutral'' torus}

From the solution of Eq. (\ref{rotde}) with $\vcirc=constant$, one can
prove that the $\alpha =4$ power-law torus is the only model of this
class characterized by a perfectly flat rotation curve.  For a generic
value of $k$, along the major axis in the edge-on projection, we have
$\vpr^2=8 k^2/(9\pi^2)$, and $\siglos^2=1/3 -8k^2/(9\pi^2)$.

A more general family of power-law tori with flat rotation curves can
be obtained by expansion of the oblate spheroid $\rho\propto 1/m^2$
(the so-called ``isothermal'' spheroid) to higher orders in the
flattening, beyond Eqs. (\ref{rhome})-(\ref{pote}). In fact, such
oblate spheroid is characterized by a flat rotation curve on the
equatorial plane (for any flattening; e.g., see BT87,
Chap. 2). Therefore, following the arguments presented in Sect. 2.1 we
find that the tori
\[
\varrho={\Rt^{2n}\over\rt^{2n+2}},\quad (n\;{\rm integer}),
\label{torn}
\]
all belong to the list of density distributions with perfectly flat
rotation curve, which includes the singular isothermal sphere and the
Mestel (1963) disk (see also Monet, Richstone \& Schechter 1981; the
toroidal solutions presented by Toomre 1982 can also be expressed as a
superposition of power-law tori). When projected face-on, much like
the isothermal sphere, the family given in Eq. (\ref{torn}) produces
the $1/R$ surface density distribution of the Mestel disk.

A torus with very unusual characteristics is the following. If we
discard one of the two homogeneous solutions of Eq. (\ref{rotde})
(corresponding to $\vcirc =0$), because of its infinite central mass,
we are left with the solution $\rhot\propto
\rt^{-5/2+\sqrt{17}/2}$. From Eqs. (20) and (\ref{suprho}) we see that
the related density is given as in Eq. (\ref{rhoT}), with
$\alpha=\alpha_n$. This torus is thus characterized by a constant
potential in the equatorial plane, so that the circular velocity
vanishes, as already pointed out in the previous Section.  From a
mathematical point of view, adding this {\it neutral} torus to a disk,
would not alter the rotation curve in the disk. In other words, if we
add the neutral torus to a flat-rotation density distribution, the
resulting density distribution will be associated with a flat rotation
curve of the same amplitude as the original.

We conclude by noting that fully self-consistent, self-gravitating
toroidal stellar systems corresponding to the density-potential pair
(\ref{rhoT})-(\ref{potT}) with $\alpha\leq\alpha_n$ do not exist.  In
fact, for $2<\alpha\leq\alpha_n$ the zero-velocity curves in the
meridional plane $(R,z)$ are open. Surprisingly, the Jeans equations
would lead to apparently ``innocent'' solutions even for
$\alpha\leq\alpha_n$.  This concrete example illustrates the risks
that are taken when stellar systems are modeled by means of the Jeans
equations without a final check on the physical viability of the
solutions found from the formal analysis.

\subsubsection{The two-integral distribution function of power-law tori}

The general considerations presented in Sect. 3.1 can be elaborated in
greater detail for the family of power-law tori.  In the case of
non-self gravitating power-law tori, the distribution function in the
presence of a dominant massive central black hole of mass $\Mbh$ can
be explicitly written as a Fricke distribution function by adopting
$h(\E)=\E^{\alpha-5/2}$ (with $\alpha>3/2$) and $f_0=3 \rhoz
a^{\alpha-2}/[2^{7/2}\pi (G\Mbh)^\alpha {\rm B}(5/2,\alpha-3/2)]$ in
Eq.(\ref{df}), where ${\rm B}(x,y)$ is the standard Euler complete
Beta function.

As noted at the end of Sect. 3.1, by combining Eqs. (\ref{rhoT}) and
(\ref{potT}), the function $\Phi=\Phi(\varrho ,\Rt^2)$ can be obtained
explicitly. In addition, for the entire family of power-law tori (with
$\alpha>\alpha_n$), the {\it envelope} and the {\it dynamical window}
(relevant to such method) can be expressed in terms of elementary
functions: for example, the envelope is defined by the relation
$\Jz^2\propto\exp (3E_c)$ when $\alpha=4$, while $\Jz^2\propto
|E_c|^{6-\alpha\over 4-\alpha}$ in all the other cases, where $E_c$ is
the energy per unit mass of a star orbiting on the circular orbit.
Finally, for two members of this family it is possible to construct
the function $\varrho =\varrho(\Phi, \Rt^2)$ explicitly, thus
providing analytically all the required ingredients for a numerical
implementation of the Hunter-Qian method (Qian et al. 1995). In
particular,
\[
\varrho=\cases{\displaystyle{
        {(\sqrt{4\Phi^2 +2\Rt^2}-2\Phi)^3\over\Rt^4},\quad (\alpha=3),}\cr
                      \displaystyle{
        {4{\rm W}^2(\Rt^2{\rm e}^{-3\Phi}/2)\over\Rt^2},\quad (\alpha=4),}
}
\]
where ${\rm W}(x)$ is the Lambert function\footnote{The Lambert
function is defined by the identity $\ln {\rm W}({\rm e}^z) + {\rm
W}({\rm e}^z)=z$.}. Remarkably, for $\alpha=3$ the density $\varrho$
can be expanded in series of $\psi^2\equiv\Rt^2/(2\Phi^2)$, and the
associated series of Fricke's terms can be re-summed in closed form as
\begin{eqnarray}
f(E,\Jz)&=&{945\over 2^{11/2}\pi}{\xi\over E^{5/2}}\left[
{_2F_1}\left({11\over 6},{13\over 6};4,-{27\xi\over 4}\right)-
         {143\xi\over 256}
{_2F_1}\left({17\over 6},{19\over 6};5,-{27\xi\over 4}\right)\right]+
                                                         \nonumber\\
        &&k\,{\rm sign}(\Jz){315\over 128\pi}{\xi\over E^{5/2}}
{_4F_3}\left({3\over 2},{5\over 2},{11\over 6},{13\over 6};
             {5\over 4},{7\over 4},4,-{27\xi\over 4}\right),
\end{eqnarray}
which is positive for $0\leq\xi\equiv\Jz^2/(4E^3)\leq 32/27$ and
$0\leq k\leq 1$ (where the upper limit refers to circular orbits and 
$_pF_q$ is the standard hypergeometric function, e.g. see Gradshteyn
\& Ryzhik 1980). Note that the odd part has been recovered under the Satoh 
ansatz, for which $\vphib =k\sqrt{\Phi}$ from Eqs. (18), (19), and
(22).  We verified by direct integration over velocity space that the
DF returns indeed the $\alpha=3$ torus.

\subsection{Scale-free oblate axisymmetric systems treated in terms of 
the Jeans equations and the Satoh decomposition}

We now apply the formulae derived in Sect. 2.1 to the family of
power-law oblate ($\epsilon=0$) spheroids. Thus, we start from
$\rho=\rhoz/m^{\gamma}$, where the range $0 <\gamma <3$ is such that
the density profile is radially decreasing and the central mass
remains finite.

Several properties of these models have been investigated by Qian et
al. (1995), who reconstructed their phase-space properties; in
addition, Evans (1994) and Evans \& de Zeeuw (1994) studied the
observational properties that would result from a decomposition
similar (but not identical) to the Satoh (1980) decomposition for a
similar class of models (which they called ``power-law'' models, in
their scale--free limit).

Here we study the relevant Jeans equations on the basis of the Satoh
decomposition, which has found widespread applications to elliptical
galaxies (e.g., see Binney, Davies \& Illingworth 1990; van der Marel,
Binney, \& Davies 1990; van der Marel 1991).  We are then able to
provide expressions for the intrinsic and projected kinematical
profiles in a form that may be useful to model the observations. These
expressions can be easily generalized to the case in which a central
black hole or a power-law dark matter halo with different flattening
are present.

From Eqs. (\ref{rhome})-(\ref{pote}) we have 
\[
\varrho={1-\gamma\eta\over\rt^{\gamma}}+
        {\gamma\eta\Rt^2\over\rt^{\gamma +2}},
\label{rhope}
\]
\[
\Phi=\cases{
     \displaystyle{-{5-\gamma -(4-\gamma)(\gamma-1)\eta\over 
                    (5-\gamma)(3-\gamma)(\gamma -2)\rt^{\gamma -2}}-
                    {\eta\Rt^2\over 
                    (5-\gamma)\rt^{\gamma}},\quad (\gamma\neq 2)},\cr
     \displaystyle{{(3-2\eta)\ln\rt\over 3}-
                   {\eta\Rt^2\over 3\rt^2},\quad (\gamma =2)},
}
\label{phipe}
\]
where the arguments described in Appendix A impose $\eta\leq
1/\gamma$; the limit $\eta =1/\gamma$ recovers the case described in
Eqs. [\ref{rhoT}]-[\ref{potT}].  The presence of a central black hole
of mass $\Mbh$ can be modeled in Eq. (\ref{phipe}) by adding a term
$-\mubh/\rt$, where $\mubh\equiv\Mbh/4\pi\rhoz a^3$.  The density in
Eq. (\ref{rhope}) becomes prominently peanut-shaped at high values of
$\eta$; this is a common property of the original model studied by 
\Sch (1979), of the flattened Plummer sphere (Lynden-Bell
1962), and of the family of ``power-law'' models (Evans 1994; the
family includes the Binney [1991] logarithmic potential as a special
case).

For $\gamma>2$ the potential vanishes at infinity and is negative
divergent at the origin.  For $0<\gamma\leq 2$ an upper truncation
must be applied to some integrals appearing in Eq. (\ref{poteI}); in
Eq. (\ref{phipe}) an additive constant has then been set to zero, so
that the potential diverges at infinity and vanishes at the origin.
Finally, for $\gamma=2$ the model is characterized by a flat rotation
curve, as the seed density distribution. In fact, 
\[
\vcirc^2 ={5-\gamma -2\eta\over (5-\gamma)(3-\gamma)\Rt^{\gamma -2}}+
           {\mubh\over\Rt}.
\label{vchom}
\]
In the entire range $0 <\gamma <3$, the models do not exhibit the
repulsive behavior noted at the end of Sect. 4.1.1 for power-law tori
and so, from this point of view, are physically acceptable.

The fully nonlinear (i.e., up to $\eta^2$ terms included) Jeans
equations (\ref{solJez})-(\ref{solJeR}) for the assumed density--potential
pair can be solved without difficulty. Here, for simplicity, we use
the solution up to first order in $\eta$; the analytical expressions
of the intrinsic and (edge-on) projected kinematical fields are given
in Appendix C. In Appendix C we also show that the projection of all
the quadratic (intrinsic) kinematical profiles can be obtained
explicitly even for $\mubh\neq 0$. Unfortunately, the streaming
velocity field cannot be obtained in explicit form when $\mubh\neq 0$;
for this case we can only provide asymptotic expressions for small
values of the projected radius. Close to the origin, independently of
the value of $\gamma$, the dynamical effects of $\mubh$ are always
dominant.

For $\mubh =0$, the analysis presented in Appendix C gives (for $3/2
<\gamma <5/2$)
\begin{eqnarray}
\Mp\,\sigap^2 &=&{\pi^{3/2}\Gamma (\gamma -3/2)\,\lt^{5-2\gamma}\over 
                 2(5-\gamma)(3-\gamma)(5-2\gamma)\Gamma (\gamma)}\times
                 \nonumber\\
              &&\left[2(5-\gamma) +\eta (2\gamma^2 -7\gamma -2)-
       {4k^2\eta\Gamma^2(3\gamma/4-1/2)\Gamma (\gamma/2)\Gamma (\gamma)\over
       \Gamma^2(3\gamma /4)\Gamma (\gamma/2-1/2)\Gamma (\gamma -3/2)}
             \right],
\end{eqnarray}
where $\lt\equiv\sqrt{\yt^2 +\zt^2}$ is the (dimensionless) circular
aperture radius and $\Gamma$ is the standard Euler gamma function.

For $\mubh\neq 0$ and $l\to 0$ (that is, in practice, for apertures
not larger than the sphere of influence of the black hole)
\begin{eqnarray}
\Mp\,\sigap^2 &=&{\mubh\pi^{3/2}\Gamma (\gamma/2 +1)\,\lt^{2-\gamma}\over 
                 2\gamma (2-\gamma)\Gamma (\gamma/2 +5/2)}\times
                 \nonumber\\
              &&\left[2(3+\gamma) - \gamma (4-\gamma)\eta -
       {k^2\eta\gamma (\gamma^2-1)\Gamma^2(\gamma +1/4)\over
       \Gamma^2(\gamma/2 +3/4)}
                \right],
\end{eqnarray}
for $1<\gamma <2$.  Thus, scale-free oblate systems with a central
black hole do have finite projected fields in the range $3/2<\gamma
<2$. In Eqs. (29)-(30) $\sigap^2$ is the {\it aperture velocity
dispersion} defined as $\sigap^2 (\ell)\equiv \int_{y^2+z^2\leq
\ell^2}\Sigma\siglos^2 dydz/\Mp (\ell)$, where $\Mp (\ell)$ is the
projected mass within $\ell$.

By combining the two formulae above we obtain an expression for the
aperture velocity dispersion which is asymptotically correct for small
and large apertures when $\mubh\neq 0$.  

\subsubsection{Effects of rotation on the aperture velocity dispersion}

In LC03, the problem of projection effects on the tilt and thickness
of the Fundamental Plane (FP) of elliptical galaxies was investigated
by means of a novel approach, introduced by Bertin, Ciotti \& Del
Principe (2002). In particular, it was shown that rotation has no
significant effects on the value of the measured ``central'' velocity
dispersion (consistent with the results obtained by van Albada, Bertin
\& Stiavelli (1995) for cuspier models), and thus has a
negligible effect on the FP tilt and thickness.  However, the models
used suffered from possessing unrealistically flat density profiles in
their central regions.  Here we illustrate how such legitimate concern
can be resolved with our method, for simplicity restricting our
discussion to the edge-on case, where the effect of rotation is
maximum.

\begin{figure}[htbp]
\includegraphics[width=0.8\textwidth]{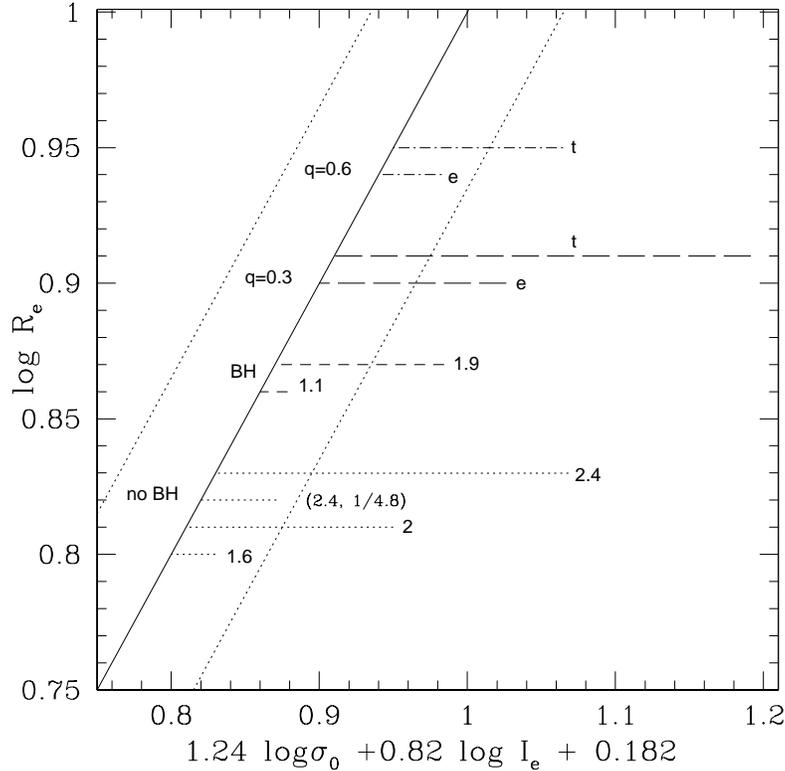}
\caption[]{Relative displacement of galaxy models induced by 
different amounts of rotational support with respect to the edge-on FP
for different slopes $\gamma$ of the density profile and different
flattenings.  The two dotted lines parallel to the FP line represent
the 1-$rms$ thickness of the FP.  The aperture velocity dispersion of
each model is normalized to that of the corresponding isotropic
rotator ($k=1$), which is placed on the FP (solid diagonal line).
Dotted and short-dashed lines (labelled by $\gamma$, lower part of the
figure) represent models of Sect. 4.2 without and with a central
dominant black hole, in the case of maximum flattening, except for the
dotted line labeled $(2.4, 1/4.8)$ which represents a model with
smaller flattening. Long-dashed and dotted-dashed lines refer to
Eq. (C11) of LC03, i.e., to the velocity dispersion of the homogeneous
oblate ellipsoid ($n=0$) calculated within an elliptical aperture with
semimajor axis equal to the effective radius $\Reff$ (label ``e'') and
over the whole model (label ``t''), for two different flattenings
(axial ratio $q=0.3$ and $q=0.6$).}
\label{fig:deltak3}
\end{figure}

In Fig. \ref{fig:deltak3} we show the position of our models with
respect to the edge-on FP. The models are first (arbitrarily) placed
on the FP (the diagonal solid line) in the isotropic case (i.e.,
$k=1$); the two dotted lines parallel to the diagonal show the 1-$rms$
dispersion of real galaxies around the FP.  We refer to the FP
coefficients measured by J{\o}rgensen, Franx \& Kj{\ae}rgaard (1996;
for Coma cluster ellipticals and $H_0= 50$ km s$^{-1}$ Mpc$^{-1}$).

The models are then ``moved'' horizontally by letting their
``observed'' velocity dispersion $\sigma_0$ vary with the Satoh
parameter $k$ down to $k=0$, while their scale-length and mean surface
brightness are kept constant.  To better illustrate our point, in
Fig. \ref{fig:deltak3} we place different models at different values
of $\Reff$.

Because of the scale-free nature of the models considered the relative
effects are independent of the adopted aperture (cfr. Eqs. [29]-[30]).
In all models $\sigma_0$ increases for decreasing $k$, i.e., as the
flattening becomes more and more supported by the azimuthal velocity
dispersion. However, the displacement depends on the density slope
$\gamma$ and on the model flattening.  Note how cuspier models are
more affected by rotation than softer ones. The effect of flattening
at fixed $\gamma$ is then shown by the line labeled $(2.4, 1/4.8)$:
for this model the density slope is $\gamma=2.4$, but the flattening
is only half of the maximum allowed.  The effect of rotation for
models with a dominant central black hole is apparently weaker.

The behavior of cuspy models, even in the presence of a massive
central black hole, is not very different from that of the Ferrers
models in LC03, thus confirming the findings of van Albada et
al. (1995) and the statistical results of LC03.

\subsection{The $v/\sigma$-ellipticity plane}

One important property of a family of models characterized by
different degrees of flattening, is the way it covers the so-called
$v/\sigma$-ellipticity plane. In fact, such plane is a simple tool
often used to investigate to what extent the observed flattening of
galaxies should be ascribed to rotation (e.g., Illingworth 1977,
Binney 1978).  For the purpose, the observed quantities ($v/\sigma$, a
measurement of the ratio of observed rotation to velocity dispersion,
and the observed ellipticity) are usually compared to a curve
associated with the behavior of classical spheroids (see Chandrasekhar
1969). However, such a comparison may be misleading, because (1)
galaxies need not conform to the properties of classical spheroids
with density stratified on homeoidal surfaces (Roberts 1962), and (2)
the observations sample the central regions and may thus provide
insufficient information for a proper comparison with the theoretical
expectations. Indeed, Evans \& de Zeeuw (1994), in the study of their
``power-law'' models, noted that their nearly isotropic models are
associated with points in the $v/\sigma$-ellipticity plane
systematically below the curve of the classical spheroids, which is
obtained from virial (and therefore integrated) quantities.

\begin{figure}[htbp]
\includegraphics[width=0.8\textwidth]{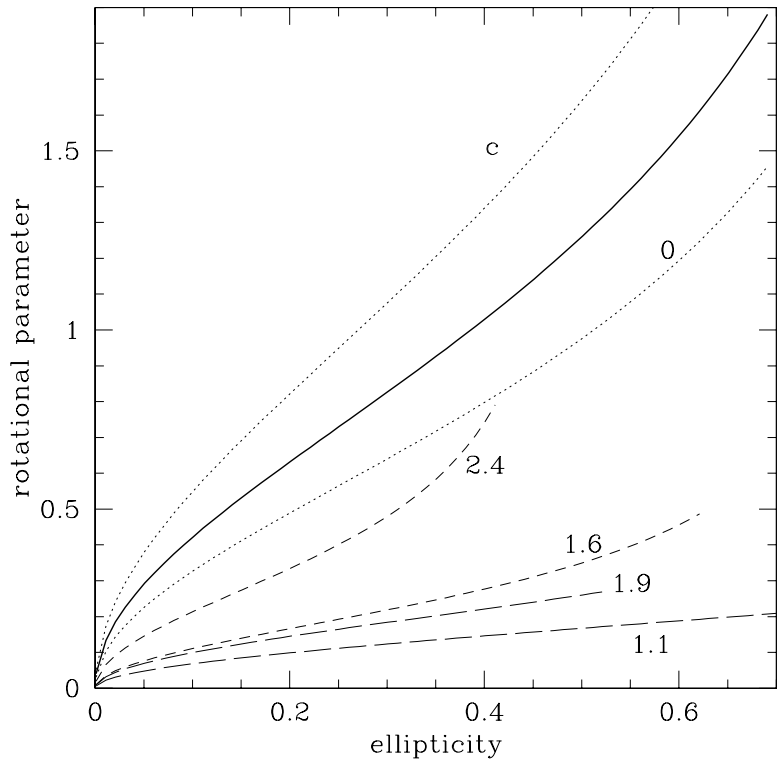}
\caption[]{The $v/\sigma$-ellipticity plane for edge-on isotropic
rotators (BT87, Eq. [4.95]).  Here the ellipticity is
defined as the value of $\eta$.  The thick solid curve is the locus of
classical (oblate) spheroids. The dotted line ``c'' refers to the
$n=0$ model in LC03, where the projected rotational velocity and the
aperture velocity dispersion are taken at the ``core'' radius and
within an elliptical aperture of semimajor axis equal to the core
radius, respectively.  The dotted line ``0'' refers to the same model,
where the projected rotational velocity is taken at $\sqrt{2/5}R_t$
(see text), while the aperture velocity dispersion is measured at the
center.  Long-dashed and short-dashed lines refer to our models, with
and without a dominant central black hole respectively, and for
different values of the density slope $\gamma$ (given by the labels) of 
$\rhoz/m^{\gamma}$.}
\label{fig:vsigma}
\end{figure}

Here we address this issue by studying the behavior of the family of
models introduced in Sect. 4.2. Note that, even though our models are
indeed obtained from a seed density distribution stratified on
homeoidal surfaces, their density distribution is not stratified on
homeoids, being the sum of a spherical density distribution and a
(negative) toroidal density. To stay closer to the issue of the
relevant observed quantities we define $\sigma$ in terms of the
aperture velocity dispersion (Eqs. [29]-[30]) and $v$ in terms of the
major axis projected streaming velocity (obtained from Eq. [C8]).
Note that, because of the scale-free nature of the present
models, the rotational parameter is independent of the aperture radius
adopted, when both the streaming velocity and the aperture velocity
dispersion are measured at the same radius.

In Fig. \ref{fig:vsigma} we show the model position in the
ellipticity-rotational parameter ($v/\sigma$) space for edge-on,
isotropic stellar systems. For the $n=0$ model described in LC03 the
major axis projected rotational velocity increases linearly with
radius and thus, following Evans \& de Zeeuw (1994) (footnote 2
there), we adopt as reference value of $v$ the value $\vpr$ measured
at the ``core'' radius (i.e. at $\sqrt{3}R_t/2$, where $R_t$ is the
ellipsoid semimajor axis). The aperture velocity dispersion $\sigma$
referred to an elliptical aperture of core radius (curve ``c''). It is
thus clear that the rotational parameter $v/\sigma$ can stay above or
below the locus of classical spheroids. In fact, it can be easily
proved that for the constant density spheroid the line-of-sight
velocity dispersion and the projected velocity dispersion coincide
(Eq. [C.10] in LC03), and so from the projected virial theorem (e.g.
Ciotti 1994) the aperture velocity dispersion over the whole object
coincides with the virial velocity dispersion (in the isotropic
case). Also, the streaming velocity, projected for the edge-on
isotropic case, evaluated along the major axis at $\sqrt{2/5}R_t$
coincides with the rotational velocity obtained from the virial
ordered kinetic energy.  For example, curve ``0'' refers to the $n=0$
model where $v$ is taken at $\sqrt{2/5}R_t$, while $\sigma$ is the
central projected velocity dispersion.

The result we wish to point out here is associated with the dashed
lines, which correspond to our models with (long-dashed) and without
(short-dashed) a dominant central black hole, for density slopes near
the limits of the acceptable range (labels near the curves give the
value of $\gamma$ for the assumed $\rho=\rhoz/m^{\gamma}$ of
Sect. 4.2). 

Thus, while our analysis confirms the importance of aperture effects
even for the classical spheroids, we are also able to show that in
general models different from classical spheroids may be found well
below the classical locus, with deviations increasing at larger
ellipticities.

\section{Concluding remarks}

The applications addressed in some detail in Sect. 4 are only a small
sample of the many applications that can grow out of the general
method introduced in this paper. Among other applications not
addressed here, we may mention the following.  MC04 adopt the method
to calculate stellar orbits in triaxial elliptical galaxies, while
Barnab\'e et al. (2005) investigate the properties of exploratory
models for the low-rotation, extraplanar gas in spiral galaxies.
Finally, we should recall that the interest in the study of toroidal
structures is also motivated by the fact that centrally depressed
surface brightness profiles need not always be associated with the
projection of spherically symmetric, non-monotonic density profiles
(Ciotti 2000), but may trace instead the presence of truly toroidal
structures.


\appendix

\section{Limits on $\epsilon$ and $\eta$}

When considering homeoidal expansions we have to impose that the
density distribution thus obtained be positive-definite, which then
sets limits on the maximum values of $\epsilon$ and $\eta$ as a
function of the seed density profile.  Without loss of generality we
take $\epsilon\leq\eta$. In the case of the expansion at fixed total
mass (see MC04, Eq. [C.1]) we require
\begin{equation}
1+\epsilon+\eta + {\epsilon y^2 +\eta z^2\over r^2}
                  {d\ln\rho (r)\over d\ln r}\geq 0.
\end{equation}
If we assume that $\rho (r)$ is a non-increasing function of $r$, by
defining $\AM\equiv\sup_{[0,\infty[}|d\ln\rho (r)/d\ln r|\geq 0$, the
inequality above is then satisfied when
\begin{equation}
\epsilon \geq (\AM -1)\eta -1,
\end{equation}
because the factor $(\epsilon y^2 +\eta z^2)/r^2$ attains its maximum 
value $\eta$ on the $z$ axis.
Clearly, if the
seed density profiles are too steep little room is left for the
expansion procedure.
In the unconstrained expansion (Eq. [\ref{rhome}]), a similar analysis
for $\epsilon\leq\eta\leq 1$ leads to the condition $1\geq\AM\eta$.

To illustrate these conclusions, let us consider the density profile
\begin{equation}
\rhot={1\over m^a(1+m^b)^c},
\label{abc} 
\end{equation}
with $a\geq 0$, $b\geq 0$, $c\geq 0$, for which $\AM = a +bc$.  Thus,
in the unconstrained expansion we have to require $\eta\leq 1/(a+bc)$,
while in the oblate case at fixed total mass the relevant condition is
$\eta\leq 1/(a+bc-1)$. Thus, for a power-law model ($c=0$),
$0\leq\eta\leq 1/a$.

A particularly interesting case of expansion at fixed total mass is
represented by the so-called $\gamma$-models (Dehnen 1993, Tremaine et
al. 1994), for which $a=\gamma$, $b=1$, $c=4-\gamma$.  The
$\gamma$-models received much attention not only for their analytical
simplicity, but also because the projected density distribution of the
$\gamma=2$ model (Jaffe 1983) is well fitted by the de Vaucouleurs
(1948) $R^{1/4}$ law. Curiously, the positivity condition for the
oblate case is $0\leq\eta\leq 1/3$, i.e, the expansion must be limited
to models rounder than E7. For the $\gamma$ models we have $\M=
1/(3-\gamma)$, and so, from Eq. (C.1) of MC04
\begin{equation}
\varrho={1+\epsilon+\eta\over\rt^{\gamma}(1+\rt)^{4-\gamma}}-
         {(\epsilon\yt^2+\eta\zt^2)(4\rt+\gamma)\over
         \rt^{2+\gamma}(1+\rt)^{5-\gamma}},
\end{equation}
with
\[
\phitz=\cases{
       \displaystyle{
       {1\over (2-\gamma)(3-\gamma)}\left[
       \left ({\rt\over 1+\rt}\right)^{2-\gamma}-1\right]
       },\quad (\gamma\neq 2);\cr
       \displaystyle{
       \ln\left({\rt\over 1+\rt}\right)
       },\quad (\gamma=2)
}
\]
\[
\phitu=\cases{
       \displaystyle{
       {1\over 3(2-\gamma)(3-\gamma)}\left[
       {\rt^{2-\gamma}(3-\gamma+\rt)\over (1+\rt)^{3-\gamma}}-1\right]-
       {\rt^{2-\gamma}F_{21}(4-\gamma,5-\gamma,6-\gamma,-\rt)\over 3(5-\gamma)}
       },\quad (\gamma\neq 2);\cr
       \displaystyle{
       {\rt-2\over 3\rt^2}+
       {1\over 3}\ln\left({\rt\over 1+\rt}\right)+
       {2\ln (1+\rt)\over 3\rt^3}
       },\quad (\gamma=2)
}
\]
\[
\phitd=\cases{
       \displaystyle{
       {F_{21}(4-\gamma,5-\gamma,6-\gamma,-\rt)\over\rt^{\gamma}(5-\gamma)}
       },\quad (\gamma\neq 2);\cr
       \displaystyle{
       {2+\rt\over\rt^4 (1+\rt)}-
       {2\ln (1+\rt)\over\rt^5}
       },\quad (\gamma=2).
}
\]
The case $\gamma=1$ was adopted by MC03 and MC04, and similar formulae
can also be obtained for the so-called modified Jaffe profile,
$\rho\propto 1/m^2 (1+m^2)$.  For $\gamma=0,1,2$ the Jeans equations
can be solved explicitly, even in the presence of a central black
hole, with functions no more complicated than standard hypergeometric
functions.

\section{Relation with expansions in spherical harmonics}

Equation (\ref{pairI}) can be recovered also from a study based on
spherical harmonics.  Let us assume that the density distribution is
given in spherical coordinates by $\rho=-\rhoz\,r\, \rhot'(r)
\cos^2\vartheta$.  In a standard expansion in spherical harmonics 
$\rho (\xv) = \sum_{l=0}^{\infty}\sum_{m=-l}^l
\Dlm(r) \Ylm(\vartheta, \varphi)$ (e.g., Jackson 1999); for our case
we thus have
\[
\Dlm (r) = -\rhoz r\rhot'(r) {\sqrt{4\pi}\over 3}\delta_{m0}
\left[\delta_{l0}+{2\over\sqrt{5}}\delta_{l2}\right].
\]
Some algebra then leads to prove that the associated potential is
indeed the one recorded in Eq. (\ref{pairI}).

In the seminal paper describing the numerical set-up of the ``\Sch
method'' for the construction of triaxial models of elliptical
galaxies (\Sch 1979), the (dimensionless) density distribution was
chosen of the form
\begin{equation}
\rhoS=F(r) - G(r){2z^2-x^2-y^2\over 2r^2}+
                  3H(r){x^2-y^2\over r^2},
\end{equation}
where $F(r) = (1+r^2)^{-3/2}$, the two non-spherical terms have
factors of the form of $l=2$ spherical harmonics, and $H$ and $G$ are
functions prescribed implicitly from separate considerations and then
calculated numerically. Given the above choice of $\rhoS$, the related
potential can be written as
\begin{equation}
\phiS=U(r) - V(r){2z^2-x^2-y^2\over 2r^2}+
                  3W(r){x^2-y^2\over r^2},
\end{equation}
and then \Sch obtained $V=V[G]$ and $W=W[H]$ by solving the
(dimensionless) Poisson equation $\nabla^2\phiS =\rhoS$ numerically.
Here we note that $V[G]$ and $W[H]$ are also available in closed
integral representation for a generic $F(r)$ (starting from
Eqs. [8]-[9] with the condition $\alpha+\beta+\gamma=0$).

Several authors constructed triaxial density-potential pairs starting
from Eqs. (B2)-(B3), through a potential-priority variation of the
\Sch method (e.g., de Zeeuw \& Merritt 1983, Hernquist \& Quinn 1989,
de Zeeuw \& Carollo 1996). We wish to emphasize that the approach
based on homeoidal expansions developed in this paper to construct
analytically tractable density-potential pairs leads to new solutions.
For example, models constructed from an expansion of triaxial
$\gamma$-models are different from those studied by de Zeeuw \&
Carollo (1996).

\section{Solutions of the Jeans equations 
         for low-flattening, scale-free axisymmetric models}

Here we record the expressions for the intrinsic and projected
kinematical profiles associated with the density profile in
Eq. (\ref{rhope}), derived from the Jeans equations under the
assumption that the underlying DF depends on $E$ and $\Jz$.  We adopt
the same normalization procedure mentioned at the beginning of
Sect. 4.

We first note that for $\gamma >1$ the edge-on projected density
(``surface brightness'') is
\[
\Sigma = {\sqrt{\pi}\Gamma (\gamma/2 -1/2)\over 
           \Gamma (\gamma/2)\lt^{\gamma-1}}
           \left[1-{(\gamma -1)\eta\zt^2\over\lt^2}\right]
\]
and the mass contained within the projected radius $\ell=\sqrt{y^2
+z^2}$ (``integrated luminosity'') is
\[
\Mp = {\pi^{3/2}(2+\eta - \gamma\eta)
        \Gamma (\gamma/2 -1/2)\over (3-\gamma)\Gamma (\gamma/2)}
        \lt^{3-\gamma}.
\]

Integration of Eq. (\ref{solJez}), truncated to first order in $\eta$,
gives
\begin{eqnarray}
\varrho\sigma^2&=&{1\over (5-\gamma)(3-\gamma)\rt^{2(\gamma-1)}}
                    \left[{5-\gamma-4\eta\over 2(\gamma -1)}-
                         {(4-\gamma)\eta\zt^2\over\rt^2}\right]+\nonumber\\
                &&{\mubh\over (3+\gamma)\rt^{1+\gamma}}\left(
                  {3+\gamma -2\gamma\eta\over 1+\gamma}-
                  {\gamma\eta\zt^2\over\rt^2}
                  \right).
\end{eqnarray}
Similarly, from Eq. (\ref{solJeR}), 
\[
\varrho(\vphidb -\sigma^2)={2\eta\Rt^2\over\rt^{2\gamma}}\left[ 
                            {1\over (5-\gamma)(3-\gamma)} + 
                            {\gamma\mubh\over (3+\gamma)\rt^{3-\gamma}}\right].
\]
Therefore, for spherical systems ($\eta =0$), the quantity above
vanishes. Finally, to leading order in $\eta$
\[
\varrho\vphib =\sqrt{\varrho\times\varrho\vphibd}\sim
               {k\sqrt{2\eta}\Rt\over\rt^{3\gamma /2}}
               \sqrt{
                     {1\over(5-\gamma)(3-\gamma)}+
                     {\gamma\mubh\over (3+\gamma)\rt^{3-\gamma}}}.
\]
The relevant projected kinematical quantities are obtained from the
formulae described in Sect. 3. Accordingly,
\[
\Sigma\Vpr^2= {2\eta k^2\sqrt{\pi}\yt^2\over\lt^{2\gamma -1}}
              \left[
              {\Gamma (\gamma-1/2)\over (5-\gamma)(3-\gamma)\Gamma(\gamma)}+
              {\gamma\mubh\over\lt^{3-\gamma}}
              {\Gamma (\gamma/2 +1)\over 2\Gamma(\gamma/2+5/2)}
              \right],
\]
and, for $\gamma>3/2$,
\begin{eqnarray}
\Sigma\sigpr^2&=&{\sqrt{\pi}\Gamma(\gamma-1/2)\over 
                  (5-\gamma)(3-\gamma)\Gamma(\gamma)\lt^{2\gamma-3}}
                  \left[{5-\gamma-4\eta\over 2\gamma-3}-
                        {(4-\gamma)\eta\zt^2\over\lt^2}\right]+\nonumber\\
               && {\mubh\sqrt{\pi}\Gamma(\gamma/2+1)\over 
                  2\Gamma(\gamma/2+5/2)\lt^{\gamma}}
                  \left({3+\gamma-2\gamma\eta\over\gamma}-
                        {\gamma\eta\zt^2\over\lt^2}\right)+
                        {1-k^2\over k^2}\Sigma\Vpr^2 .
\end{eqnarray}
Unfortunately, when $\mubh\neq 0$ no explicit expression is available
for the quantities $\Sigma\vpr$ and $\Sigma\vpr^2$. This latter
quantity is obtained by expanding to lowest order in $\eta$ the
function $(\Sigma\vpr)^2/\Sigma$. In two interesting cases, namely
when $\mubh=0$ and when $\mubh\neq 0$ and $\ell\to 0$, asymptotic
expressions can be obtained from the projection integral by
substituting $\xt^2=\rt^2-\lt^2$ and then by changing the integration
variable to $\rt/\lt$. One then finds
\begin{equation}
\Sigma\vpr  =-\cases{
             \displaystyle{
             {\sqrt{2\pi\eta}k\yt\over\lt^{3\gamma/2 -1}}
              {\Gamma(3\gamma/4-1/2)\over
              \sqrt{(5-\gamma)(3-\gamma)}\Gamma(3\gamma /4)},
             \quad (\mubh =0);
             }\cr
             \displaystyle{
             {\sqrt{2\pi\eta\mubh}k\yt\over\lt^{\gamma +1/2}}
              {\sqrt{\gamma}\Gamma(\gamma/2+1/4)\over
              \sqrt{3+\gamma}\Gamma(\gamma/2+3/4)},
              \quad (\mubh\neq 0,\; \lt\to 0)},
}
\end{equation}
and 
\begin{equation}
\Sigma\vpr^2=\cases{
             \displaystyle{{2\sqrt{\pi}\eta k^2\yt^2\over\lt^{2\gamma -1}}
             {\Gamma^2(3\gamma/4-1/2)\Gamma (\gamma/2)\over
              (5-\gamma)(3-\gamma)\Gamma^2(3\gamma /4)\Gamma (\gamma/2-1/2)},
             \quad (\mubh =0);}\cr
             \displaystyle{
             {2\sqrt{\pi}\eta\mubh k^2 \yt^2\over\lt^{2+\gamma}}
             {\gamma\Gamma^2(\gamma/2+1/4)\Gamma (\gamma/2)\over
             (3+\gamma)\Gamma^2(\gamma/2+3/4)\Gamma (\gamma/2-1/2)},
             \quad (\mubh\neq 0,\;\lt\to 0)}.
}
\end{equation}
Near the origin, the black hole contribution to the various intrinsic
and projected fields dominates for any value of $\gamma$.

\end{document}